\documentclass{ws-procs975x65}

\begin{document}

\title{INTERACTING DARK ENERGY AND TRANSIENT ACCELERATED EXPANSION}

\author{W. ZIMDAHL}

\address{Universidade Federal do Esp\'{\i}rito Santo,\\
Vit\'{o}ria, ES, Brazil\\
E-mail: winfried.zimdahl@pq.cnpq.br}

\author{C.Z. VARGAS}

\address{Centro Brasileiro de
Pesquisas F\'{\i}sicas -- CBPF,\\
Rio de Janeiro, Brazil\\
E-mail: czuniga@cbpf.br}

\author{W. S. HIP\'{O}LITO-RICALDI}

\address{Universidade Federal do Esp\'{\i}rito Santo,\\
S\~ao Mateus, ES, Brazil\\
E-mail: hipolito@ceunes.ufes.br}

\begin{abstract}
We argue that interactions in the dark sector may have a crucial impact on the cosmological dynamics.
In particular, the future cosmic evolution may be very different from that predicted by the $\Lambda$CDM model. An example is a scenario in which the currently observed accelerated expansion is an interaction-induced  transient phenomenon. We discuss such type of behavior on the basis of a two-fluid toy model.
\end{abstract}


\bodymatter

\section{Introduction}\label{intro}
Although the $\Lambda$CDM model has received the status of a standard model for the cosmological dynamics,
alternative approaches are useful to open different perspectives and therefore continue to deserve attention.
We are focusing here on the occasionally discussed idea that the currently observed accelerated expansion of the Universe might be a transient phenomenon, implying that a de Sitter phase does not necessarily represent the  final state of the cosmic evolution. While naturally a different future evolution is not directly testable with
current data, the latter may nevertheless be used to restrict the admissible range of model parameters.
Within an effective fluid approach we establish a toy model in which a suitable interaction between dark matter and dark energy results in a transition from decelerated to accelerated expansion in the past, followed by a reverse transition from accelerated to decelerated expansion in the future\cite{transient,cristofher}.

\section{Background Dynamics of Interacting Fluids}

\begin{figure}[!h]
\centering
\ \\
\ \\
\includegraphics[width=0.43\textwidth]{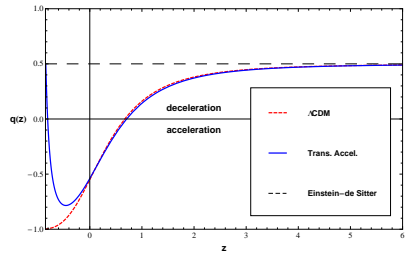}
\caption{The deceleration parameter of the transient acceleration model as function of the redshift
for the best-fit parameters (solid line). The dashed line shows the corresponding dependence for the $\Lambda$CDM model. The value $q= 1/2$ corresponds to the Einstein-de Sitter universe.}
\label{figq}
\end{figure}

We model the cosmic substratum as a two-component system, described by the energy-momentum tensor
\begin{equation}\label{split}
T^{ik} = T_{m}^{ik} + T_{x}^{ik}\ , \qquad T_{\ ;k}^{ik} = 0\ ,
\end{equation}
where $T_{m}^{ik}$ represents pressureless (dark) matter and $T_{x}^{ik}$ represents some kind of dark energy.
Both the total energy momentum tensor $T^{ik}$ and its parts $T_{m}^{ik}$ and $T_{x}^{ik}$ are
assumed to have
perfect-fluid structures, i.e., $T^{ik} = \rho u^{i}u^{k} + p h^{ik}$ and
$T_{A}^{ik} = \rho_{A} u_A^{i} u^{k}_{A} + p_{A} h_{A}^{ik}$, respectively, with $h^{ik} = g^{ik} + u^{i} u^{k}$ and  $ h_{A}^{ik} = g^{ik} + u_A^{i} u^{k}_{A}$, where  $A = m, x$.
In general, the four-velocities $u^{i}$, $u^{i}_{m}$ and $u^{i}_{x}$ are different. We assume them to coincide, however, in the homogeneous and isotropic background, where
\begin{equation}\label{balfundo}
\dot{\rho}_{m} + 3H \rho_{m} = Q \ ,\quad \dot{\rho}_{x} +
3H (1+w)\rho_{x} = - Q \ ,
\end{equation}
with $Q = - u_{a}Q^{a}$ being a phenomenological quantity.
An interaction modifies the usual $a^{-3}$ behavior of the
matter energy density to
$\rho_{m} =\rho_{m0}a^{-3}\,f(a)$,
where the function $f(a)$ encodes the influence of the interaction and $Q = \rho_{m}\dot{f}/f$.
It is convenient to write the function $f(a)$ as
$f\left(a\right)=1+g\left(a\right)$.
We consider the special case
\begin{equation}\label{gauss}
w= -1 \ , \quad g(a)=\gamma\, a^5 \exp (-a^2/\sigma ^2)\ ,
\end{equation}
where $\gamma$ is an interaction constant, as a toy model.
Under these circumstances the dark-energy density
becomes
\begin{equation}\label{rhoxtrans}
\rho_{x} = \rho_{x_{0}}^{eff}
- \gamma\,\frac{\rho_{m_0}\left[a^{2} - \frac{3}{2}\sigma^{2}\right]}{1+g_{0}}\,\,e^{-a^{2}/\sigma^{2}},\quad
\rho_{x_{0}}^{eff}
= \rho_{x_{0}} - \frac{3}{2}\gamma\,\frac{\rho_{m_0}\left[\sigma^{2} - \frac{2}{3}\right]}{1+g_{0}}
e^{-1/\sigma^{2}},
\end{equation}
where $\rho_{x_{0}}^{eff}$ is an effective cosmological constant.
The interaction re-normalizes the bare (interaction-free) value $\rho_{x_{0}}$.
A transient acceleration is only possible for $\rho_{x_{0}}^{eff}=0$ since otherwise the constant
would always prevail in the long-time limit. This means, for accelerated expansion to be a transient phenomenon, part of the interaction has to cancel the bare cosmological constant.
With the requirement $\rho_{x_{0}}^{eff} = 0$ and the definition
\begin{equation}\label{defK}
 K = \frac{8 \pi G}{3 H_{0}^{2}}\gamma\frac{\rho_{m_0}}{1+g_{0}}
\end{equation}
of the interaction parameter $K$ one finds that there exists a range
\begin{equation}\label{range}
\frac{2}{9}\frac{e^{1/\sigma^{2}}}{\sigma^{2} - \frac{2}{3}} < K <
\frac{2 e^{1/\sigma^{2}}}{3\sigma^{2}}
\end{equation}
of possible values for the interaction parameter $K$ that guarantees an early matter-dominated phase, accelerated expansion around the present time and a phase of decelerated expansion in the far-future limit.
The corresponding deceleration parameter is shown Fig.~\ref{figq}. Using the Constitution set, we find best-fit values $K = 0.018$ and $\sigma = 5.23$.

\section{Perturbation Dynamics}

Different from the $\Lambda$CDM model, in dynamical dark-energy models perturbations of the dark-energy component do not vanish in general.
In order to reduce the first-order gauge-invariant perturbation dynamics we assume a proportionality
$\delta_{x} = \epsilon\delta_{m}$ between the fractional, gauge-invariant perturbations $\delta_{x}$ and $\delta_{m}$ of dark energy and dark matter, respectively.
Calculating the matter power-spectrum and comparing the result with the data of the 2dFGRS project, we find a best-fit value of
$\epsilon = - 0.000023$ which indicates that for our model dark-energy perturbations are negligible, at least on scales that are relevant for structure formation. However, our analysis also reveals that on the largest scales a value of $\epsilon = 0.001$ shows a better performance. This indicates an increasing role of  dark-energy perturbations with increasing scale, even though they remain small compared with the matter perturbations.  A more advanced analysis should take into account a scale-dependence of the parameter $\epsilon$.

\section{Discussion}
A phenomenological model of transient accelerated expansion in which an interaction in the dark sector is constitutive for the cosmological dynamics does not seem to contradict current observational data.
The detailed structure of the interaction was chosen here for mathematical convenience but we think that it can be used to discuss general features of transient acceleration models.
The interaction has to play a twofold role. It has both to cancel a ``bare" cosmological constant and, at the same time, to generate a phase of accelerated expansion by itself.
A statistical  analysis on the basis of the 2dFGRS data revealed that perturbations of the dark-energy component
are negligible on scales that are relevant for structure formation.
Only on very large scales their contribution might be noticeable.

\section*{Acknowledgments}
Financial support by CAPES and CNPq is gratefully acknowledged.



\begin{thebibliography}{99}
\bibitem{transient} J.C. Fabris, B. Fraga, N. Pinto-Neto and W. Zimdahl,
{\em JCAP} {\bf 1004}, 008 (2010).
\bibitem{cristofher}
C. Zu\~niga Vargas, W.S. Hip\'{o}lito-Ricaldi and W. Zimdahl, {\em JCAP} {\bf 1204}, 032 (2012).

\end{thebibliography}
\end{document}